\documentclass[final]{aipproc}
\layoutstyle{8x11double}

\begin{document}

\title{Gauged $U(1)_R$ supergravity on orbifold 
\footnote{
Talk presented at the 11th International 
Symposium on Particles, Strings and Cosmology, 
May 30 - June 4, 2005, Gyeongju, Korea}}

\classification{04.65.+e, 12.60.Jv}
\keywords{Supergravity, Supersymmetric models}

\author{Hiroyuki Abe\footnote{E-mail address: 
abe@gauge.scphys.kyoto-u.ac.jp}}{
address={Department of Physics, Kyoto University, Kyoto 606-8502, Japan}}

\begin{abstract}
We discuss a gauged $U(1)_R$ supergravity on five-dimensional 
orbifold ($S^1/Z_2$) in which a $Z_2$-even $U(1)$ gauge field 
takes part in the $U(1)_R$ gauging, and show the structure of 
Fayet-Iliopoulos (FI) terms allowed in such model. 
Some physical consequences of the FI terms are examined. 
\end{abstract}

\maketitle

\section{Introduction}
Recently five-dimensional (5D) supergravity (SUGRA) on the orbifold 
$S^1/Z_2$  has been studied as an interesting theoretical framework for 
physics  beyond the SM. It has been noted that 5D orbifold SUGRA with a 
$U(1)_R$ symmetry gauged by the $Z_2$-odd graviphoton can provide the 
supersymmetric Randall-Sundrum (RS) model~\cite{Gherghetta:2000qt} 
in which the weak to Planck scale hierarchy can arise naturally from 
the geometric localization of 4D graviton~\cite{Randall:1999ee}, and/or 
Yukawa hierarchy can be generated by the quasi-localization of the 
matter zero modes in extra dimension where we generically have an 
interesting correlation between the flavor structure in the sparticle 
spectra and the hierarchical Yukawa couplings~\cite{Choi:2003di}. 
In the former case, the bulk cosmological constant and brane tensions 
which are required to generate the necessary AdS$_5$ geometry appear 
in the Lagrangian as a consequence of the $U(1)_R$ FI term with 
$Z_2$-odd coefficient. 

In this talk we consider a more generic orbifold SUGRA which contains 
a $Z_2$-even 5D gauge field $A^X_\mu$ participating in the $U(1)_R$ 
gauging~\cite{Abe:2004nx}. If 4D $N=1$ SUSY is preserved by the 
compactification, the 4D effective theory of such model will contain 
a gauged $U(1)_R$ symmetry associated with the zero mode of $A^X_\mu$, 
which is not the case when the 5D $U(1)_R$ is gauged only through the 
$Z_2$-odd graviphoton. Based on the known off-shell 
formulation~\cite{Fujita:2001bd}, we formulate a gauged $U(1)_R$ SUGRA 
on $S^1/Z_2$ in which both $A^X_\mu$ and the graviphoton take part in 
the $U(1)_R$ gauging and then analyze the structure of FI terms allowed 
in such model. As expected, introducing a $Z_2$-even $U(1)_R$ gauge field 
accompanies new bulk and  boundary FI terms in addition to the known 
integrable boundary FI term which could be present in the absence of any 
gauged $U(1)_R$ symmetry~\cite{Ghilencea:2001bw}. As we will see, those new 
FI terms can have interesting implications to the quasi-localization of the 
matter zero modes in extra dimension and the SUSY breaking~\cite{Choi:2003di} 
and also to the radion stabilization.

\section{Formulation}
For a minimal setup, we introduce two vector multiplets 
and two hypermultiplets in the off-shell formulation of 
5D (conformal) SUGRA~\cite{Fujita:2001bd}: 
\begin{eqnarray}
\begin{array}{rcl}
{\cal V}_Z &=&
\big( M^Z=\alpha,\, A^Z_\mu,\, \Omega^{Zi},\, Y^{Zij} \big), \\
{\cal V}_X &=&
\big( M^X=\beta,\, A^X_\mu,\, \Omega^{Xi},\, Y^{Xij} \big), 
\end{array}
\nonumber
\end{eqnarray}
and 
\begin{eqnarray}
{\cal H}_c &=&({\cal A}^x_i,\eta^x,{{\cal F}}^x_i), 
\nonumber \\
{\cal H}_p &=& (\Phi^x_i,\zeta^x, F^x_i), 
\nonumber
\end{eqnarray}
with the norm function 
\begin{eqnarray}
{\cal N} &=& \alpha^3 -\frac{1}{2} \alpha \beta^2, 
\nonumber
\end{eqnarray}
and the hypermultiplet gauging 
\begin{eqnarray}
\Big( {t}_Z,\,{t}_X \Big)\Phi &=&
\Big( c\epsilon(y),\, q \Big)i\sigma_3\Phi, 
\nonumber \\
\Big( {t}_Z,\,{t}_X \Big){\cal A}&=&
\Big( \textstyle{-\frac{3}{2} k \epsilon(y)}, 
-r \Big)i\sigma_3{\cal A},  
\nonumber
\end{eqnarray}
where we adopt the $2 \times 2$ matrix notations 
omitting $x=1,2$ index and $SU(2)$ indices $i,j=1,2$, 
and the hyperscalars satisfy the reality condition 
${\cal A}^\ast=i\sigma_2 {\cal A} i\sigma_2^T$, 
$\Phi^\ast=i\sigma_2 \Phi i\sigma_2^T$. 
The $Z_2$-even bosonic (non-auxiliary) components 
are $\alpha$, $A^Z_y$, $A^X_{\underline\mu}$, 
${\cal A}^{x=2}_{i=2}$ and $\Phi^{x=2}_{i=2}$, and 
${\cal V}_Z$, ${\cal H}_c$ are the graviphoton vector 
multiplet and the compensator hypermultiplet respectively. 
The $Z_2$-odd coefficient $\epsilon(y)$ 
in the hypermultiplet gauging is consistently introduced 
by the mechanism proposed in~\cite{Bergshoeff:2000zn}. 
The nonzero value of the charge $r$ corresponds to 
the $U(1)_R$ symmetry gauged by $Z_2$-even vector field 
$A^X_{\underline\mu}$. 

The bosonic part of the Lagrangian is given by 
\begin{eqnarray}
{\cal L}_{\rm bosonic}
&=& {\cal L}_{\rm bulk}
+{\cal L}_{\partial \epsilon}
+{\cal L}_{N=1}, 
\nonumber
\end{eqnarray}
\begin{eqnarray}
e^{-1}{\cal L}_{\rm bulk} 
&=& 
-{\textstyle \frac{1}{2}}R
-{\textstyle \frac{1}{4}}a_{IJ}
F^I_{\mu \nu} F^{\mu \nu J} 
+{\textstyle \frac{1}{2}}a_{IJ} 
\nabla^m M^I \nabla_m M^J 
\nonumber \\ &&
+{\textstyle \frac{1}{8}}e^{-1} C_{IJK} 
\epsilon^{\lambda \mu \nu \rho \sigma} A^I_\lambda 
 F^J_{\mu \nu} F^K_{\rho \sigma} 
\nonumber \\ && 
+{\rm tr} \Big[ |\nabla_m \Phi|^2
-|\nabla_m {\cal A}|^2-|V_{m}|^2 
\nonumber \\ && 
-M^IM^J \big( 
\Phi^\dagger {t}_I^\dagger {t}_J \Phi 
-{\cal A}^\dagger {t}_I^\dagger {t}_J {\cal A} 
\big) \Big] 
\nonumber \\ &&
-{\textstyle \frac{1}{2}} {\rm tr}\Big[ 
{\cal N}_{IJ} Y^{I\dagger}Y^J
-4Y^{I\dagger} 
\big( {\cal A}^\dagger {t}_I {\cal A} 
-\Phi^\dagger {t}_I \Phi \big)\Big], 
\nonumber \\
e_{_{(4)}}^{-1}{\cal L}_{\partial \epsilon} 
&=& 
-2\alpha \big( 3k 
+{\textstyle \frac{3}{2}}k\, 
{\rm tr} \left[ \Phi^\dagger \Phi \right] 
+c\, {\rm tr} \left[ \Phi^\dagger \sigma_3 \Phi 
\sigma_3 \right] \big) 
\nonumber \\ && \qquad 
\times \left( \delta(y)-\delta(y-\pi R) \right), 
\nonumber \\
e_{_{(4)}}^{-1} {\cal L}_{N=1} 
&=& 
M_{_{(4)}}^2 \Big[ \, 
-2r \Big(\, 2Y^{X(3)}-e^{-1}e_{_{(4)}}\partial_y\beta \, \Big) 
-{\textstyle \frac{1}{2}}R^{(4)} \Big] 
\nonumber \\ && \qquad 
\times \left( \Lambda_0 \delta(y) 
+\Lambda_\pi \delta(y-\pi R) \right), 
\nonumber
\end{eqnarray}
where the matrix notations are employed again, 
$I,J=(Z,X)$, $a_{IJ}=-\frac{1}{2} 
\frac{\partial^2\ln {\cal N}} {\partial M^I\partial M^J}$, 
$M_{(4)}^2=\big( 1 +\frac{1}{2}
{\rm tr} \big[ \Phi^\dagger \Phi \big] \big)^{2/3}$ and 
$V_m = \frac{1}{2} \big( \Phi^\dagger (\nabla_m \Phi) 
-(\nabla_m \Phi)^\dagger \Phi \big)
-\frac{1}{2} \big( {\cal A}^\dagger (\nabla_m {\cal A}) 
- (\nabla_m {\cal A})^\dagger {\cal A} \big)$. 
Here we have included only 4D $N=1$ {\it pure} SUGRA 
action at the orbifold fixed points without any 
K\"ahler and superpotentials for simplicity. 
We remark that after the superconformal gauge fixing, 
\begin{eqnarray}
{\cal N}=1, && 
{\cal A}=\mathbf{1}_2 \sqrt{1+{\rm tr}[\Phi^\dagger \Phi]/2}, 
\nonumber
\end{eqnarray}
we find the bulk FI term 
$-e \big( 6k Y^{Z\,(3)} +4r Y^{X\,(3)} \big)$ 
in ${\cal L}_{\rm bulk}$ and the boundary FI term 
$-2r\, e_{(4)} M_{(4)}^2 \big(2Y^{X\,(3)}
-e^{-1}e_{(4)} \partial_y \beta \big) 
\big( \Lambda_0 \delta(y)+\Lambda_\pi \delta(y-\pi R) \big)$ 
in ${\cal L}_{N=1}$ for the auxiliary fields $Y^{Z,X}$ 
in the vector multiplets. 

We are interested in the 4D Poincar\'e invariant 
background geometry, 
\begin{eqnarray}
ds^2=e^{2K(y)} \eta_{\underline\mu \underline\nu} 
dx^{\underline\mu} dx^{\underline\nu}-dy^2, 
\nonumber
\end{eqnarray}
and the gravitino-, hyperino- and gaugino-Killing 
parameters on this background are given respectively by 
\begin{eqnarray}
\kappa &=& \partial_y K - {\cal P}/3 
\nonumber \\
F &=& \partial_y v - (q \beta + c\epsilon(y) \alpha-{\cal P}/2)v 
\nonumber \\
D &=& \partial_y \phi + g^{\phi \phi}{\cal P}_\phi 
\nonumber \\ && \qquad 
-2r M_{(4)}^2 g^{\phi \phi} \beta_\phi 
\big( \Lambda_0 \delta(y)-\Lambda_\pi \delta(y-\pi R) \big), 
\nonumber
\end{eqnarray}
where
\begin{eqnarray}
{\cal P} &=& 
-2 \Big[ {\textstyle \frac{3}{2}}k\epsilon(y)\alpha +r \beta 
\nonumber \\ && \qquad 
+\left\{ \left( {\textstyle \frac{3}{2}}k+c \right) \epsilon(y) \alpha 
+(r+q)\beta \right\} v^2 \Big], 
\nonumber
\end{eqnarray}
and $\phi$ is the physical gauge scalar field parameterizing 
the (very special) manifold of vector multiplet determined 
by ${\cal N}=\alpha^3(\phi)-\alpha(\phi) \beta^2(\phi)/2=1$ 
with the metric $g_{\phi \phi}=a_{IJ}M^I_\phi M^J_\phi$. 
We choose $\alpha(\phi)=\cosh^{2/3}(\phi)$ and 
$\beta(\phi)=\sqrt{2}\cosh^{2/3}(\phi)\tanh(\phi)$ 
in the following. 
The real and diagonal component of the quaternionic hyperscalar 
field $\Phi$ is represented by $v$ in the Killing parameters, 
and zero vacuum values are assumed for the other components 
for simplicity. In terms of these Killing parameters, the 4D 
energy density is found to be 
\begin{eqnarray}
E &=& \int \,dy \,\, e^{4K} 
\Big( \frac{1}{2} g_{\phi\phi}D^2 
+\frac{2}{1+v^2}|F|^2 -6|\kappa|^2 \Big), 
\nonumber
\end{eqnarray}
and it is obvious that the Killing condition 
$\kappa=D=F=0$ determines a stationary point 
of the 4D scalar potential if the solution exists.

\section{Physical consequences}
Now we examine some physical consequences of the 
5D gauged $U(1)_R$ supergravity on $S^1/Z_2$ 
which can have the bulk and the boundary FI term, 
for the supersymmetric vacuum configurations, 
$\kappa=D=F=0$. 

First we consider the case that we have a charged 
hypermultiplet $\Phi$ with the charge satisfying $q/r<-1$. 
For $k=c=0$ that results in $K(y) \simeq 0$, the 
vacuum values of the scalar fields are given by 
\begin{eqnarray}
\phi=0, && 
v=v_0 \equiv 
\pm \textstyle{\sqrt{-\frac{r}{r+q}}}, 
\nonumber
\end{eqnarray}
for $\Lambda_{0,\pi}=0$, and 
\begin{eqnarray}
\phi &\simeq& 
-2\textstyle{\frac{|rv_0|}{rv_0}\sqrt{1+\frac{r}{q}}}
(A_+ e^{\omega y} -A_- e^{-\omega y}), 
\nonumber \\
v &\simeq& v_0 + (A_+ e^{\omega y} +A_- e^{-\omega y}), 
\nonumber
\end{eqnarray}
for $\Lambda_{0,\pi} \ne 0$, where 
\begin{eqnarray}
A_\pm &=& 
\pm \frac{\sqrt{2}|rv_0|}{2v_0}
\left( \frac{q}{r+q} \right)^{7/6} 
\frac{\Lambda_0+\Lambda_\pi e^{\pm \omega \pi R}}
{e^{\pm 2 \omega \pi R}-1}, 
\nonumber
\end{eqnarray}
and $\omega=\sqrt{-8rq}$. 
We find a nontrivial $y$-dependent vacuum values 
for the latter case due to the boundary FI term. 
Notice that the vacuum value of the gauge scalar 
$\phi(y)$ gives the $y$-dependent 
mass for the charged hypermultiplets which results 
in nontrivial zero-mode wavefunctions for them. 
We will show the zero-mode profile in the next 
more simple but interesting case. 

\begin{figure}
\begin{minipage}{1.5\linewidth}
\begin{minipage}{0.48\linewidth}
   $\phi(y)$\\
   \centerline{\includegraphics[width=\linewidth]{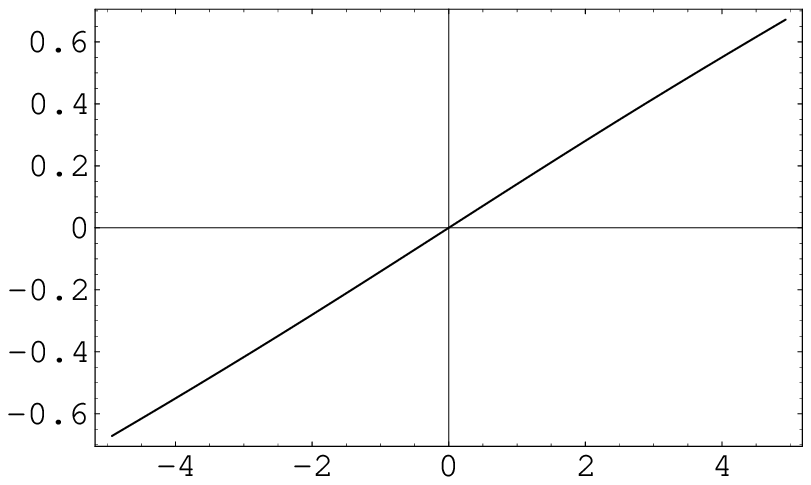}$y$}
\end{minipage}
\hfill
\begin{minipage}{0.48\linewidth}
   $\phi(y)$\\
   \centerline{\includegraphics[width=\linewidth]{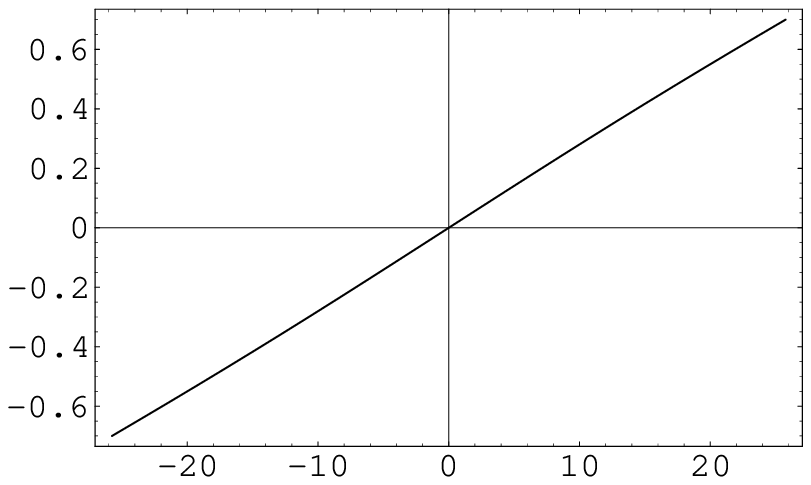}$y$}
\end{minipage}\\*[15pt]
\begin{minipage}{0.48\linewidth}
   $\log_{10} \Phi^{(0)}(y)/\Phi^{(0)}(\pi R)$\\
   \centerline{\includegraphics[width=\linewidth]{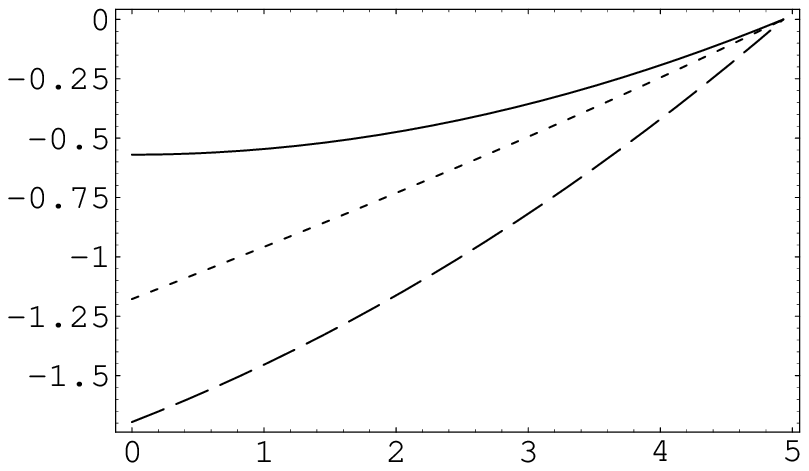}$y$}
\end{minipage}
\hfill
\begin{minipage}{0.48\linewidth}
   $\log_{10} \Phi^{(0)}(y)/\Phi^{(0)}(\pi R)$\\
   \centerline{\includegraphics[width=\linewidth]{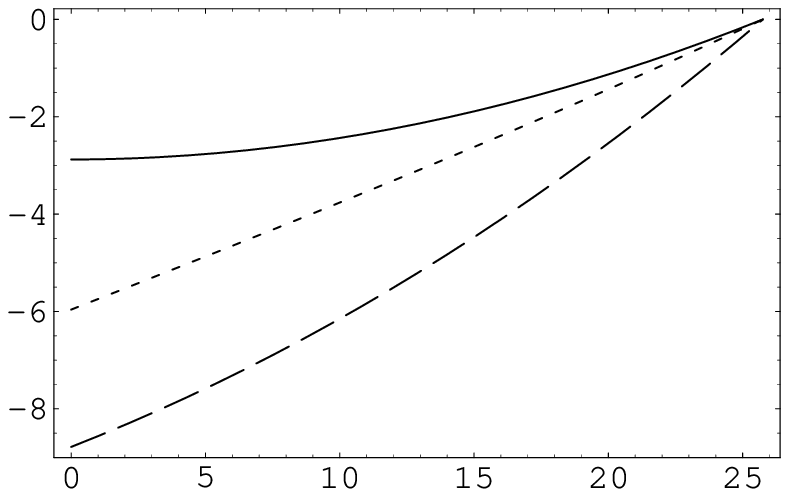}$y$}
\end{minipage}\\*[15pt]
\begin{minipage}{0.48\linewidth}
   \centerline{(a)\ $r=0.05$,\ $k=0$}
\end{minipage}
\hfill
\begin{minipage}{0.48\linewidth}
   \centerline{(b)\ $r=0.01$,\ $k=0$}
\end{minipage}
\end{minipage}
\caption{The profiles of $\phi$ and the matter zero mode 
$\Phi^{(0)}$ for some cases with $k=0$ and $\lambda_0=0$.
Here we choose $\lambda_\pi=(r-1)/2$. 
For the matter zero mode profile, the solid-,
dotted- and dashed-curves represent the case
with $(q,c)=(0.5,0)$, $(0,0.5)$ and $(0.5,0.5)$, respectively.
All the curves are shown within $|y| \le \pi R$.}
\label{fig:1}
\end{figure}

\begin{figure}
\begin{minipage}{1.5\linewidth}
\begin{minipage}{0.48\linewidth}
   $\phi(y)$\\
   \centerline{\includegraphics[width=\linewidth]{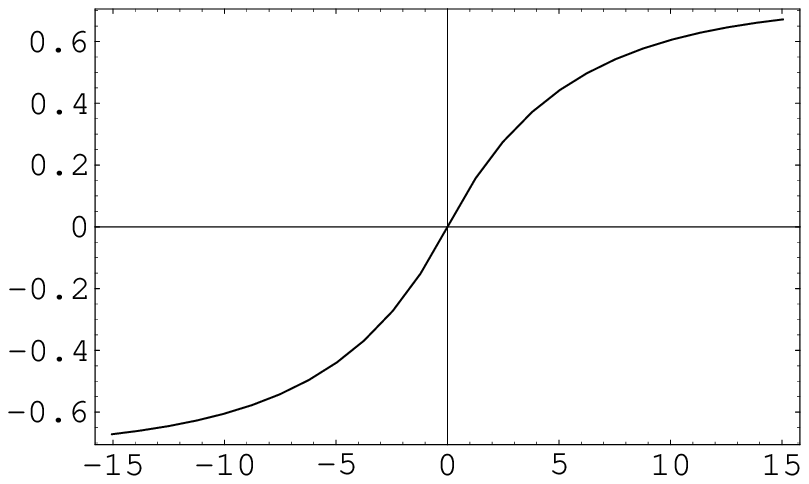}$y$}
\end{minipage}
\hfill
\begin{minipage}{0.48\linewidth}
   $\phi(y)$\\
   \centerline{\includegraphics[width=\linewidth]{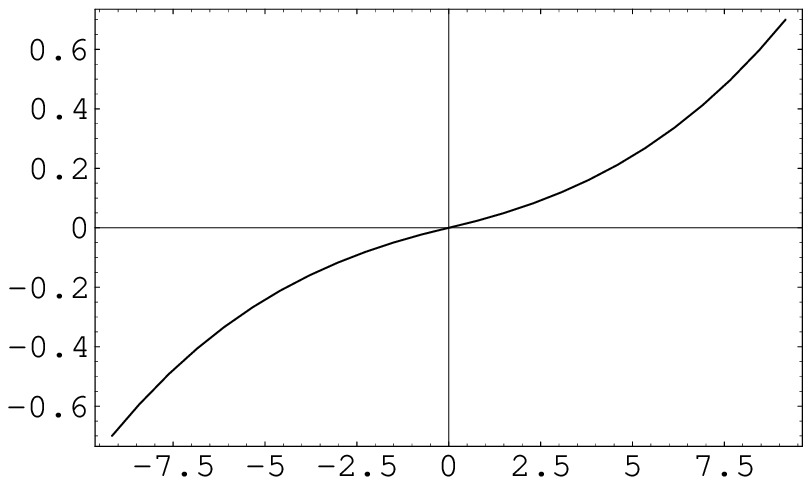}$y$}
\end{minipage}\\*[15pt]
\begin{minipage}{0.48\linewidth}
   $\log_{10} \Phi^{(0)}(y)/\Phi^{(0)}(\pi R)$\\
   \centerline{\includegraphics[width=\linewidth]{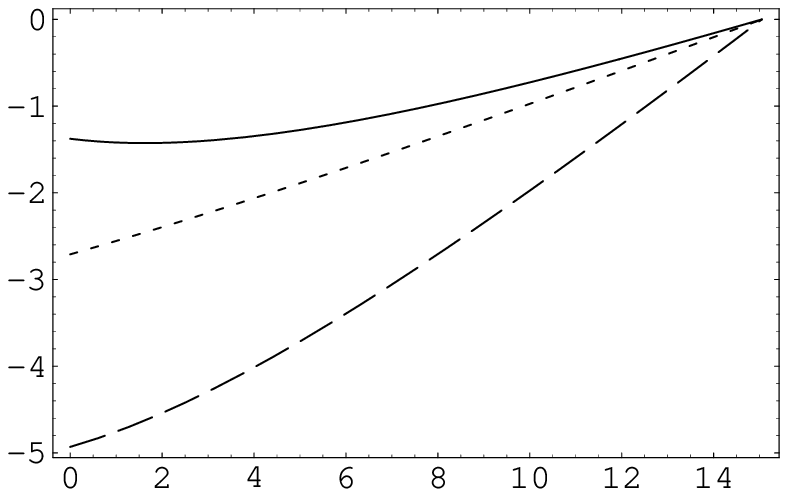}$y$}
\end{minipage}
\hfill
\begin{minipage}{0.48\linewidth}
   $\log_{10} \Phi^{(0)}(y)/\Phi^{(0)}(\pi R)$\\
   \centerline{\includegraphics[width=\linewidth]{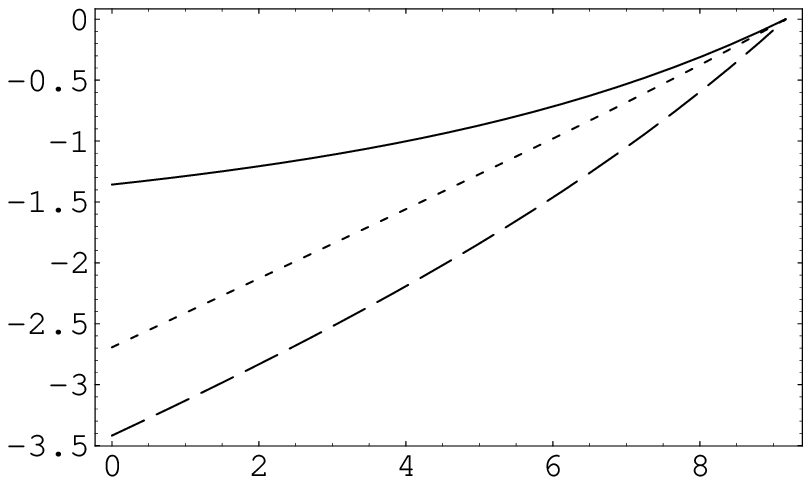}$y$}
\end{minipage}\\*[15pt]
\begin{minipage}{0.48\linewidth}
   \centerline{(a)\ $r=0.05$,\ $k=-0.1$}
\end{minipage}
\hfill
\begin{minipage}{0.48\linewidth}
   \centerline{(b)\ $r=0.01$,\ $k=0.1$}
\end{minipage}
\end{minipage}
\caption{The profiles of $\phi$ and $\Phi^{(0)}$
for $r,k \ne 0$, $\lambda_0=0$ and $\lambda_\pi=(r-1)/2$.
Again the solid-, dotted- and dashed-curves represent the case
$(q,c)=(0.5,0)$, $(0,0.5)$ and $(0.5,0.5)$, respectively.
Note that $K \simeq -ky$ in this supersymmetric solution.}
\label{fig:2}
\end{figure}

Next we consider the case there are charged chiral 
multiplets $Z_{0,\pi}$ with the charge $q^z_{0,\pi}$ 
at the orbifold fixed points $y=0,\pi R$ respectively, 
but no hypermultiplets with the charge $q/r<-1$ in bulk. 
We introduce minimal K\"ahler potential and no 
superpotential for them at the fixed points. 
For $k=c=0$, the vacuum values of the scalar fields 
are given by 
\begin{eqnarray}
\phi=2\sqrt{2}ry+\sqrt{2}\lambda_0, && v=0, 
\nonumber
\end{eqnarray}
where 
$\lambda_{0,\pi}=(r +\sum_z q^z_{0,\pi} 
|z_{0,\pi}|^2) \Lambda_{0,\pi}$ and the orbifold 
radius is determined by 
$2\pi R=-\frac{\lambda_0+\lambda_\pi}{r}$. 
We find a linear profile of $\phi$ in the $y$-direction 
due to the bulk FI term, which results in the Gaussian 
form of the zero-mode wavefunction for the charged 
hypermultiplet, 
\begin{eqnarray}
\Phi^{(0)}(y) 
&\simeq& \Phi^{(0)}(0) e^{2(q+r)(ry^2+\lambda_0 y)}. 
\nonumber
\end{eqnarray}
The ratio of the wavefunction values between two fixed 
points are then shown to be 
$\frac{\Phi^{(0)}(0)}{\Phi^{(0)}(\pi R)} 
\approx e^{-\frac{q+r}{2r}(\lambda_\pi^2-\lambda_0^2)}$. 
Some numerical plots are shown in Fig.~\ref{fig:1} 
for $c \ne 0$ but $k=0$ and in Fig.~\ref{fig:2} 
for both $c,k \ne 0$. 
From these figures we find that the nonvanishing $r$ 
(i.e., gauging $U(1)_R$ by $Z_2$-even vector field) 
as well as the bare kink mass $c$ affects the zero-mode 
profiles of the charged hypermultiplets significantly. 
The nonvanishing charge $k$ changes the linear profile 
of $\phi$ resulting in a more/less severe localization 
of the charged hypermultiplet zero-mode, depending on the 
sign of $kr$. 

\section{Summary}
We have studied  a 5D gauged $U(1)_R$ supergravity on 
$S^1/Z_2$ in which both a $Z_2$-even $U(1)$ gauge field 
and the $Z_2$-odd graviphoton take part in the $U(1)_R$ gauging. 
Based on the off-shell 5D supergravity of Ref.~\cite{Fujita:2001bd}, 
we examined the structure of Fayet-Iliopoulos (FI) terms allowed by 
such theory. As expected, introducing a $Z_2$-even $U(1)_R$ gauging 
accompanies new bulk and  boundary FI terms in addition to 
the known integrable boundary FI term which could be present 
in the absence of any gauged $U(1)_R$ symmetry. 
The new (non-integrable) boundary FI terms  originate from the $N=1$ 
boundary supergravity, and thus are free from the bulk supergravity 
structure in contrast to the integrable boundary FI term which is 
determined by the bulk structure of 5D supergravity~\cite{Ghilencea:2001bw}.

We have examined some physical consequences of the $Z_2$-even $U(1)_R$ 
gauging in several simple cases. It is noted that the FI terms of gauged 
$Z_2$-even $U(1)_R$  can lead to an interesting deformation of vacuum 
structure which can affect the quasi-localization of the matter zero 
modes in extra dimension and also the SUSY breaking and radion 
stabilization. Thus the 5D gauged $U(1)_R$ supergravity on orbifold has 
a rich theoretical structure which may be useful for understanding some 
problems in particle physics such as the Yukawa hierarchy and/or the 
supersymmetry breaking~\cite{Choi:2003di}. For such phenomenological study 
and for the analysis of the radion stabilization, the $N=1$ superfield 
description~\cite{PaccettiCorreia:2004ri} will be useful. 
When one tries to construct a realistic particle physics model within 
gauged $U(1)_R$ supergravity, one of the most severe constraint will 
come from the anomaly cancellation condition. 
In some cases the Green-Schwarz mechanism might be necessary to 
cancel the anomaly, which may introduce another type of FI term 
into the theory~\cite{Dudas:2004ni}. 
These issues will be studied in future works.

\begin{theacknowledgments}
The author would like to thank Kiwoon~Choi for the 
collaboration~\cite{Abe:2004nx} which forms the 
basis of this talk. This work was supported by 
KRF PBRG 2002-070-C00022. 
\end{theacknowledgments}

\end{document}